\documentclass[conference]{IEEEtran}
\IEEEoverridecommandlockouts
\usepackage{cite}
\usepackage{amsmath,amssymb,amsfonts}
\usepackage{algorithmic}
\usepackage{graphicx}
\usepackage{textcomp}
\usepackage{xcolor}
\usepackage{url}
\usepackage[caption=false]{subfig}
\usepackage{multirow}
\usepackage{tabularx}
\usepackage{makecell}
\usepackage{caption}
\usepackage{xcolor,colortbl}
\usepackage[ruled,vlined,linesnumbered]{algorithm2e}
\definecolor{g1}{RGB}{200, 200, 200}

\usepackage{tikz}
\usetikzlibrary{fit,calc}

\makeatletter
\def\thickhline{%
  \noalign{\ifnum0=`}\fi\hrule \@height \thickarrayrulewidth \futurelet
   \reserved@a\@xthickhline}
\def\@xthickhline{\ifx\reserved@a\thickhline
               \vskip\doublerulesep
               \vskip-\thickarrayrulewidth
             \fi
      \ifnum0=`{\fi}}
\makeatother

\newlength{\thickarrayrulewidth}
\colorlet{gray1}{gray!60}

\usepackage{soul}

\newcommand{\ignore}[1]{}

\def\BibTeX{{\rm B\kern-.05em{\sc i\kern-.025em b}\kern-.08em
    T\kern-.1667em\lower.7ex\hbox{E}\kern-.125emX}}
\begin{document}

\title{HySec-Flow: Privacy-Preserving Genomic Computing with SGX-based Big-Data Analytics Framework}

\author{
\resizebox{0.55\linewidth}{!}{
\begin{tabular}{cccc}
Chathura Widanage$^1$ & Weijie Liu$^1$ & Jiayu Li$^1$ & Hongbo Chen$^1$\\
  XiaoFeng Wang$^2$ &Haixu Tang$^2$ & Judy Fox$^{3}$ \\
\multicolumn{4}{c}{$^{1,2}$Indiana University} \\
\multicolumn{4}{c}{$^{3}$University of Virginia} \\

\multicolumn{4}{c}{$^1$\{cdwidana,weijliu,jl145,hc50\}@iu.edu}\\ \multicolumn{4}{c}{$^2$\{xw7,hatang\}@indiana.edu} \\  \multicolumn{4}{c}{$^3$\{ckw9mp\}@virginia.edu}

\end{tabular}}\\
}

\maketitle
% Page number
%\thispagestyle{plain}
%\pagestyle{plain}
%============
\begin{abstract}
Trusted execution environments (TEE) such as Intel's Software Guard Extension (SGX) have been widely studied to boost security and privacy protection for the computation of sensitive data such as human genomics. However, a performance hurdle is often generated by SGX, especially from the small enclave memory. In this paper, we propose a new Hybrid Secured Flow framework (called "HySec-Flow") for large-scale genomic data analysis using SGX platforms. Here, the data-intensive computing tasks can be partitioned into independent subtasks to be deployed into distinct secured and non-secured containers, therefore allowing for parallel execution while alleviating the limited size of Page Cache (EPC) memory in each enclave.
We illustrate our contributions using a workflow supporting indexing, alignment, dispatching, and merging the execution of SGX- enabled containers. We provide details regarding the architecture of the trusted and untrusted components and the underlying Scorn and Graphene support as generic shielding execution frameworks to port legacy code. We thoroughly evaluate the performance of our privacy-preserving reads mapping algorithm using real human genome sequencing data. The results demonstrate that the performance is enhanced by partitioning the time-consuming genomic computation into subtasks compared to the conventional execution of the data-intensive reads mapping algorithm in an enclave. The proposed HySec-Flow framework is made available as an open-source and adapted to the data-parallel computation of other large-scale genomic tasks requiring security and scalable computational resources.

\end{abstract}

\begin{IEEEkeywords}
Privacy-preserving Computing; Software Guard Extension (SGX); Reads mapping.
\end{IEEEkeywords}
\section{Introduction}

Security and privacy issues have received increasing attention in big-data analytics performed on public or commercial clouds. In particular, personal genomic data contain identifiable information concerning human individuals: it has been shown that the identity of a participant in a human genome study could be revealed from her genetic profile through searching online genealogy databases~\cite{gymrek2013identifying}. As a result, biomedical researchers are cautious of moving the intensive computation involving personal human genomic data onto the commercial cloud.

Cryptographic techniques are available to protect data privacy on the cloud.  {\em Homomorphic encryption} (HE) ~\cite{Fontaine2007} allows users to perform computation directly on encrypted data. However, HE introduces several magnitudes of computational overheads. A promising alternative has recently been presented by a new generation of hardware supporting a \textit{trusted execution environment} (TEE), in which sensitive data are kept on secure storage and processed in an isolated environment, called the \textit{enclave}.  A prominent example is the Intel Software Guard Extension (SGX)~\cite{anati2013innovative}, which has a set of instructions for establishment and management of an enclave on Intel's mainstream processors, which are available in major cloud service providers such as Microsoft Azure ~\cite{russinovich2017introducing}. Current benchmark experiments on data-intensive computing tasks\cite{shaon2017sgx} demonstrate that SGX provides data protection against attacks from the host operating system or even system administrators while introducing only moderate computation overhead; therefore, it is widely considered to be suitable for data-intensive computation, including the computing tasks involving personal human genomic data.

\begin{figure}[ht]
\centering
\includegraphics[width= 0.85\linewidth]{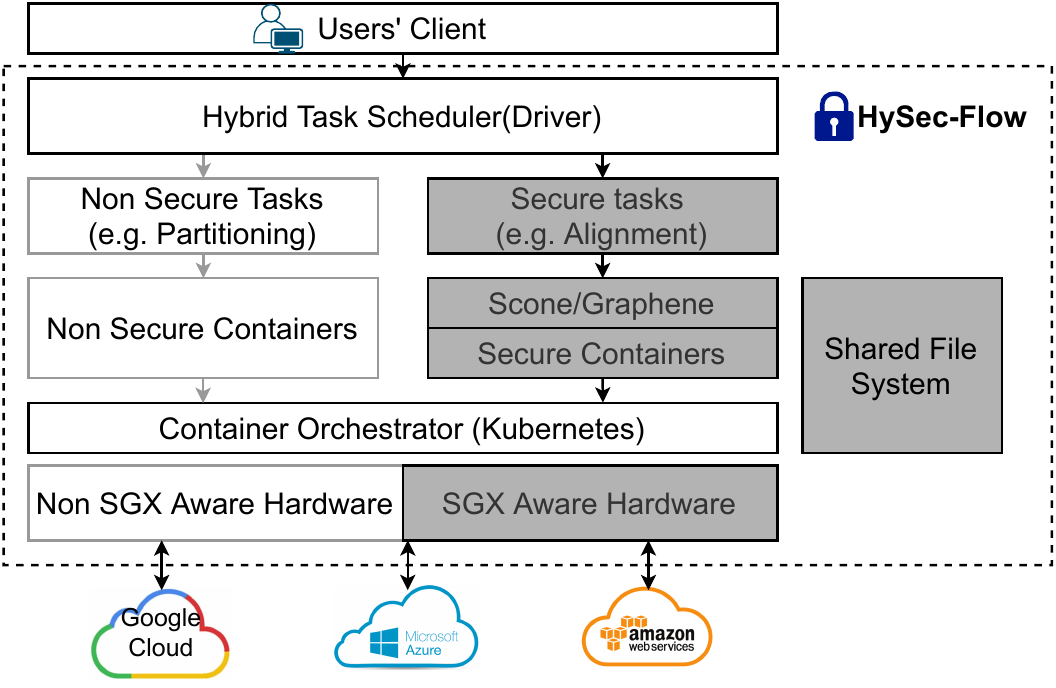}
\caption{Framework Overview.}\label{fig:framework-overview}

\end{figure}

Privacy-preserving algorithms have been developed for several genomic analysis tasks, including genetic testing and variant searching using human genomic profiles ~\cite{chen2017presage, chen2017princess, carpov2018secure}. These tasks are relatively lightweight and do not require extensive memory that exceeds the limited Page Cache (EPC) available in an enclave. Hence, the efforts of these implementations were focused on the data encryption/decryption and the protection of the data from side-channel information leaks (e.g., using data oblivious algorithms ~\cite{mandal2018data}). More recently, privacy-preserving algorithms ~\cite{ kockan2020sketching, pascoal2021dyps} were developed for Genome-wide Association Studies (GWAS), a common computational approach to identifying the associations between phenotypes and genetic variants ~\cite{tam2019benefits}. These methods exploited sketching algorithms to reduce the memory usage of GWAS computation to be executed inside the enclave within the limits of EPC memory. However, the sketching algorithms were customized for the specific computing task (i.e., GWAS) and cannot be generalized to other tasks. Furthermore, privacy-preserving approaches are still lacking for parallel data-intensive computation using multiple enclaves enabled by SGX.

\begin{figure}[ht]
\vspace{-1ex}
  \centering
    \includegraphics[width=0.9\linewidth]{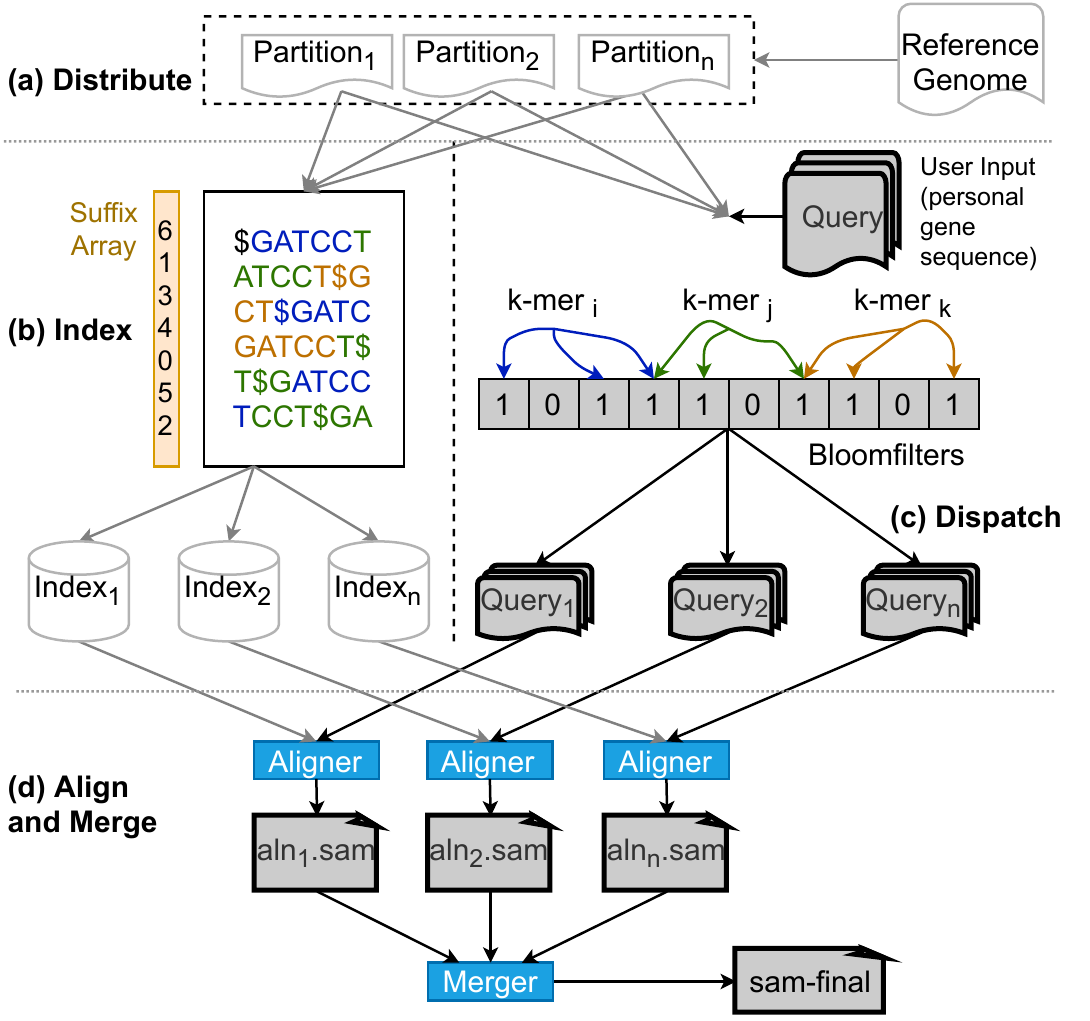}
  \caption{The conventional (Untrusted) workflow: the data with gray background needs to be encrypted because it involves private information.}
  \label{fig:workflow}
\vspace{-2ex}
\end{figure}
This paper presents a generic privacy-preserving data analytics framework for developing large-scale genome computing tasks using SGX.  A key challenge here is that only limited resources are directly accessible by the code running inside the enclave. Therefore, it is critical to devise a sophisticated method to partition the target computation task into subtasks so that each subtask can be executed efficiently using the enclave when necessary. 

It is worth noting that HySec-Flow distinguishes itself from the current approaches (e.g., Scone~\cite{arnautov2016scone} and Graphene-SGX~\cite{tsai2017graphene}) that provide runtime environments to run existing programs inside the enclave. As shown in our benchmark experiments, these approaches do not support either a hybrid enclave/non-enclave architecture or the use of parallel computation with multiple enclaves and so are not scalable for large data-intensive genome computing tasks. Furthermore, the efficiency of running specific algorithms is not optimized inside the enclave in the original applications. As well as subtasking, other  contributions of the paper include:
\begin{itemize}
\item We design a hybrid task scheduler to integrate secure and non-secure containers into HySec-Flow for performing the subtasks in enclaves to address scaling issues of SGX.
\item We demonstrate the design strategy of our analytics framework using the implementation of the {\em reads mapping} task (i.e., the alignment of millions of short ( $\approx100$ bases long) DNA sequences ({\em reads}) acquired from a human individual onto a reference human genome).
\end{itemize}

Reads mapping serves as a prerequisite step for many downstream analyses in human genome computing (e.g., genome variation calling, genotyping, gene expression analysis), and thus many software tools (e.g., BWA~\cite{li2009fast}, Bowtie~\cite{langmead2012fast}) have been developed for this fundamental task. Notably, the reads acquired from a human individual contain identifiable information about the donor and should be protected in a public cloud environment. Previously, customized algorithms were proposed for privacy-preserving reads mapping using cryptography approaches~\cite{chen2012large}, which introduced significant computing overheads and did not scale well with massive demands.

To the best of our knowledge, HySec-Flow is the first SGX-based privacy-preserving solution of reads mapping that introduced reasonable computing overhead while is highly parallelizable and scalable. Our novel hybrid task scheduler with secure containers enables a workflow for complex analysis such as a modified reads mapping and alignment algorithm. The end-to-end secure analysis framework, as shown in Fig. 1 is released as open-source software at \cite{code_framework}.

\section{Background}
\subsection{Intel SGX}
Intel SGX is a set of x86 instruction extensions that offer hardware-based memory encryption and isolation for application code and data. The protected memory area (called an \textit{enclave}) resides in an application's address space, providing confidentiality and integrity protection. SGX is a user-space TEE characterized by flexible process-level isolation: a program component can get into an enclave mode and be protected by execution isolation, memory encryption, and data sealing against the threats from the untrusted OS and processes running in other enclaves. More specifically, the memory of an enclave is mapped to a special physical memory area called Enclave Page Cache (EPC). It is encrypted by Memory Encryption Engine (MEE) and cannot be directly accessed by other system software.
Such protection, however, comes with in-enclave resource constraints. Particularly, only 128 MB (256 MB for some new processors) encryption-protected memory (called Enclave Page Cache or EPC) is reserved. Although virtual memory support is available, it incurs significant overheads in paging.

\subsection{SGX-based Data Sealing}

SGX remote attestation allows a remote user to verify that the enclave is correctly constructed and runs on a genuine SGX platform. In Intel's attestation model, three parties are involved: (1) The Independent Software Vendor (ISV) who is registered to Intel as the enclave developer; (2) The Intel Attestation Service (IAS) hosted by Intel which verifies the enclave; and (3) The SGX platform, which operates the SGX enclaves. The attestation begins with the ISV sending an attestation request challenge, which can be generated by an enclave user who wants to perform the attestation of the enclave. The attested enclave then generates a verification report including the enclave measurement, which can be verified by an Intel-signed quoting enclave (QE) through \textit{local attestation}. The QE signs the report using the attestation key, and the generated \textit{quote} is forwarded to the Intel Attestation Service (IAS).  The IAS verifies the quote and signs the verification result using the Intel private key. The verification result can convince the ISV or the enclave user by verifying the signature and comparing the enclave measurement.
When an enclave is instantiated, it protects the data by keeping it within the enclave boundary. In general, the secrets provisioned to an enclave are lost when the enclave is closed. However, if the private data must be preserved during one of these events for future use within an enclave, it must be stored outside the enclave boundary before closing the enclave. To protect and preserve the data, a mechanism is in place which allows enclave software to retrieve a key unique to that enclave that the enclave can only generate on that particular platform. Using that key, the enclave software can encrypt data, store them on the platform, or decrypt the encrypted data are stored on the platform. SGX refers to these encryption and decryption operations as {\em sealing} and {\em unsealing}, respectively.
When data needs to be encrypted and stored outside the enclave,  {\em sealing} and {\em unsealing} are needed.
Using sealing, the data within the enclave is encrypted using an encryption key derived from the CPU hardware. 

Intel SGX provides two policies for encryption keys: MRENCLAVE (enclave identity) and MRSIGNER (signing identity). These policies affect the derivation of the encryption key and are described in the documentation of Intel SGX ~\cite{costan2016intel}. 
Developers can take advantage of sealing based on the Signing Identity policy to share sensitive data via a sealed data blob between multiple enclaves initiated by a single application and/or those by different applications. 
To utilize Intel SGX's data sealing feature, we use the set of keys generated and stored in the processor's fuse array.
There are two identities associated with an enclave. The first is the Enclave Identity and is represented by the value of MRENCLAVE, which is a cryptographic hash of the enclave log (measurement) as it goes through every step of the build and initialization process. MRENCLAVE uniquely identifies any particular enclave, so using the Enclave Identity will restrict access to the sealed data only to instances of that enclave. Therefore, we use the other key policy provided by SGX - MRSIGNER, which generates a key based on the value of the enclave's MRSIGNER and the enclave's version. Specifically, we encapsulate the \verb|sgx_seal_data()| function, to better leverage the key derived from the instruction \texttt{EGETKEY}. We also implement utility codes for encrypting the initial genome data.  
\subsection{Bloom filter}
\label{sec:bloomfilter}
A Bloom filter is a space-efficient probabilistic data structure. It provides membership queries over dynamic sets with an allowable false positive rate. %In bioinformatics, the Bloom filter has been recently utilized in applications such as k-mer counting, genome assembly and contamination detection\cite{chikhi2013space,melsted2011efficient,salikhov2014using,stranneheim2010classification,chu2014biobloom}.
%In bioinformatics, the Bloom filter has been utilized in applications such as k-mer counting, genome assembly and contamination detection\cite{salikhov2014using}.
\begin{figure}[ht]
\vspace{-1ex}
\centering
\includegraphics[width=0.56\linewidth]{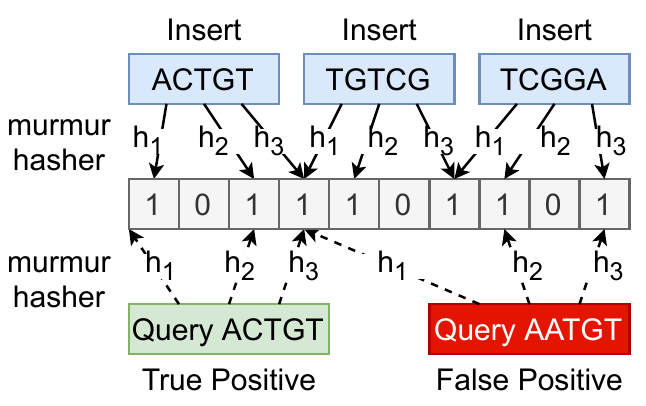}
\caption{Conventional Bloom filter with $k=3$ that illustrates the true positive, and false positive.}
\label{fig:bloomfilter}
\vspace{-1ex}
\end{figure}
An ordinary Bloom filter consists of a bit array $B$ of $m$ bits, which are initially set to $0$, and $k$ hash functions, $h_1, h_2, . . ., h_k$, mapping keys from a universe $U$ to the bit array range $\{1, 2, . . .,m\}$. 
In order to insert an element $x$ from a set $S = \{x_1, x_2, . . ., x_n\}$into the filter, the bits at positions $h_1(x), h_2(x), . . ., h_k(x)$ are set to $1$. To query if an element q is in the filter, all of the bits at positions $h_1(q), h_2(q), . . ., h_k(q)$ are examined. 
If at least one bit is equal to $0$, then $q$ is not in $S$. 
Otherwise, $q$ likely belongs to the set. %The uncertainty stems from coincidental cases, where all the corresponding bits, $h_i(q) = 1, 2, . . ., k$, may be equal to one even though $q$ is not in $S$. Such a chance occurrence is called a false positive hit, and its probability is called the 
The false positive rate 
%The probability for a false positive hit depends on the selection of the parameters $m$ and $k$, the size of the bit array, and the number of hash functions, respectively. After inserting n distinct elements at random to the bit array of size m, the probability that a specific bit in the filter is $0$ is $(1-1/m)^{kn}$. 
%Therefore, the false positive rate is:
$F = (1-(1-\frac{1}{m})^{kn})^k \approx (1-\exp{(-k/r)})^k,$
where $r = m/n$ is the number of bits per element. %Minimizing the equation for a fixed ratio of $r$ yields the optimal number of hash functions of $k = r\log{2}$, in which case the false positive rate is $(0.6185)^r$ \cite{broder2004network}.

\subsection{Threat Model}

For HySec-Flow, we follow the classical SGX threat model. Denial-of-Service (DoS), side-channel attacks, and physical attacks against the CPU are out of scope~\cite{lee2017inferring,wang2017leaky} and can be tackled by different techniques (e.g., mitigating the negative effect of Hyper-threading~\cite{chen2018racing,oleksenko2018varys}). Similarly, enclaves are trusted and free of vulnerabilities.

\section{Architecture}

\begin{figure*}[ht]
  \centering
    \includegraphics[width=0.8\linewidth,height = 0.26\linewidth]{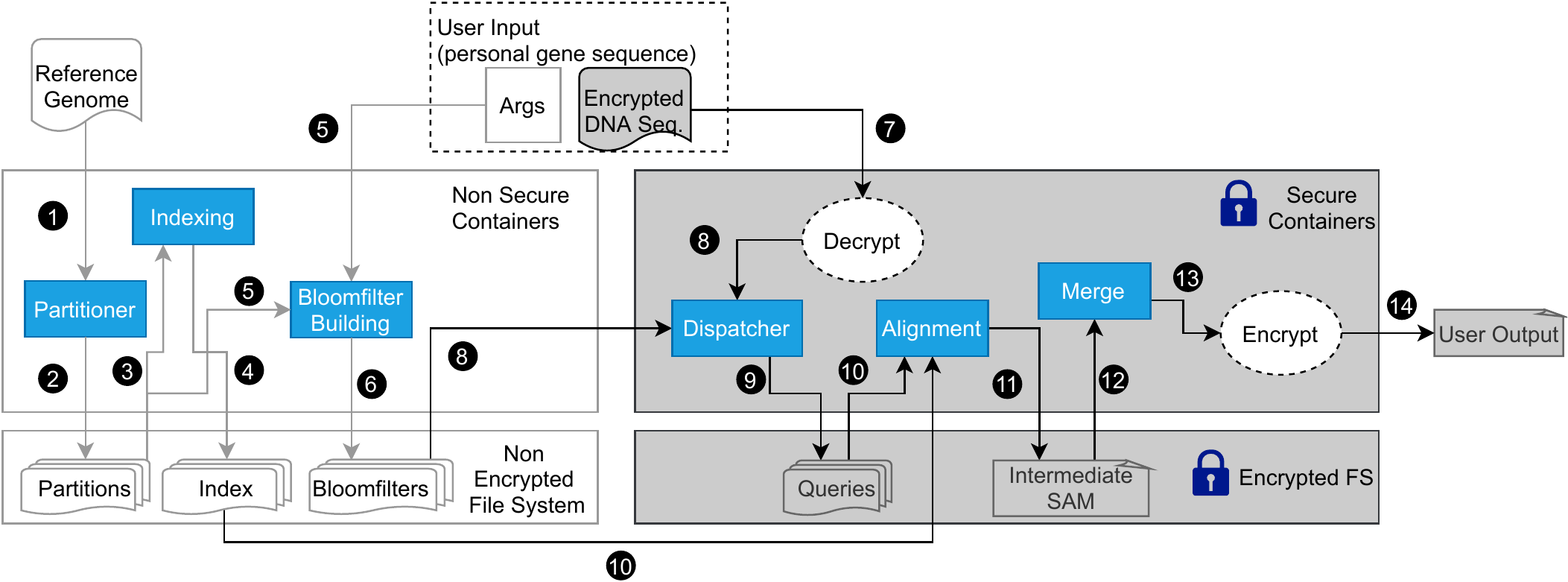}
  \caption{The Workflow of Privacy-Preserving Genomic Computing Framework with Hybrid Containers and Resources.}
\label{fig:overall_overview}
\vspace{-3ex}
\end{figure*}

In the framework shown in Fig. \ref{fig:framework-overview}, the driver (task scheduler) runs on the central control node where a computing task first splits into subtasks, and the subtasks are then deployed to worker nodes for execution. Subtasks are deployed with APIs of existing orchestration tools on their distributed platforms (e.g., Kubernetes and Docker Swarm) and the framework-specific communication mechanism between the workers (secure/non-secure containers). In this paper, we implement the reads mapping algorithm using the proposed framework.
\begin{algorithm}[t]
\small
\caption{Distributed Pipeline}
\label{alg:Pipeline}
\SetKwInOut{Input}{input}
\SetKwInOut{Output}{output}
\SetKwProg{Fn}{Function}{:}{}
\SetKwFunction{GetUniqueBloomFilterName}{GetUniqueBloomFilterName}
\SetKwFunction{GenerateBloomFilters}{GenerateBloomFilters}
\SetKwFunction{DISPATCH}{DISPATCH}
\SetKwFunction{ALIGNMENT}{ALIGNMENT}
\SetKwFunction{WriteToDisk}{WriteToDisk}
\SetKwFunction{ReadBFFromDisk}{ReadBFFromDisk}
\SetKwFunction{WriteSAMToDisk}{WriteSAMToDisk}
\SetKwFunction{ReadQuery}{ReadQuery}
\SetKwFunction{MERGE}{MERGE}
\SetKwFunction{PIPLINE}{PIPLINE}
\SetKwFunction{ReadSAMs}{ReadSAMs}
\SetKwFunction{WriteResultToDisk}{WriteResultToDisk}
\SetKwFunction{merge}{merge}
\Input{$G = \{g_1,g_2,\dots ,g_p\}$ : Reference genome partitions; \\
$I$: Input DNA Sequence;\\
$args$: Program arguments}
\Output{$Q = \{q_1,q_2,\dots ,q_p\}$ : Partitioned user input; \\
$M = \{m_1,m_2,\dots ,m_p\}$ : Reads mapping per partition;\\
$S$ : Final output\\}

\Fn{\PIPLINE{G, I, args}}
{
B = \GenerateBloomFilters{G, args}\\
\colorbox{gray!30}{Q = []} \tcp*[h]{protected fs}\\
\For { p in $\{ 1\dots G.length \}$ distributed }
{
\colorbox{gray!30}{Q[p] = \DISPATCH{B[i], I, args}} \tcp*[h]{protected}\\
}

\colorbox{gray!30}{M = []} \tcp*[h]{protected fs}\\
\For { p in $\{ 1\dots G.length \}$ distributed }
{

\colorbox{gray!30}{M[p] = \ALIGNMENT{G[p], Q[p]}} \tcp*[h]{protected}

}

\colorbox{gray!30}{S = \MERGE{M}}\tcp*[h]{protected}

}
\vspace{-1ex}
\end{algorithm}

We've devised a workflow-based approach to address the privacy issues in the existing read mapping software tools while adding the capability to parallelize the computation workload across a cluster of Intel SGX-enabled nodes. The DIDA framework\cite{mohamadi2015dida} for parallel execution inspires our implementation of reads mapping on high-performance computing platforms while designing the SGX-based implementation for the subtasks involving sensitive data (i.e., input reads). It first partitions the reference genome into multiple segments and then uses bloom filters to partition the input set of reads into subsets, which is assigned to a segment that the reads are likely mapped onto. In the next step, multiple subtasks are deployed and each substask involves a segment mapping to a subset of reads. Notably, although Intel SGX provides hardware-assisted encryption and protection of sensitive data, it comes with a performance cost due to the limited size of the EPC (Enclave Page Cache) available for the computation inside the enclave. Hence, we have to be careful only to move data that needs to be protected into the SGX enclaves and perform only the computations involving sensitive data inside the SGX enclaves. This clear separation helps to minimize the data that needs to be moved into the protected space and minimize the computation overhead introduced by EPC swaps. 

Fig.~\ref{fig:overall_overview} shows the detailed workflow of our implementation for SGX-based secure reads mapping. We have adapted four significant tasks from the DIDA framework that will be executed to perform genome sequencing. The driver node accepts the job, while the worker nodes are used for the data pre-processing and read alignment. The partitioning of the reference genome into segments and their indexing is a one-time process. Furthermore, the reference genome is public and such a step can be performed without using SGX.
\begin{algorithm}[t]
\small
\caption{Internal operations within the framework}
\label{alg:functions}
\SetKwInOut{Input}{input}
\SetKwInOut{Output}{output}
\SetKwProg{Fn}{Function}{:}{}
\SetKwFunction{GetUniqueBloomFilterName}{GetUniqueBloomFilterName}
\SetKwFunction{GenerateBloomFilters}{GenerateBloomFilters}
\SetKwFunction{DISPATCH}{DISPATCH}
\SetKwFunction{ALIGNMENT}{ALIGNMENT}
\SetKwFunction{WriteToDisk}{WriteToDisk}
\SetKwFunction{ReadBFFromDisk}{ReadBFFromDisk}
\SetKwFunction{WriteSAMToDisk}{WriteSAMToDisk}
\SetKwFunction{ReadQuery}{ReadQuery}
\SetKwFunction{MERGE}{MERGE}
\SetKwFunction{PIPLINE}{PIPLINE}
\SetKwFunction{ReadSAMs}{ReadSAMs}
\SetKwFunction{WriteResultToDisk}{WriteResultToDisk}
\SetKwFunction{merge}{merge}
\Input{$G = \{g_1,g_2,\dots ,g_p\}$ : Reference genome partitions; \\
$I$: Input DNA Sequence}

\Fn{\DISPATCH{b, I, args}}
{
$q = []$\\

\tcp*[h]{reading sequences of the input}\\  
\For {seq in I}{ 
  		\For {bmer in seq}
  		{	\If{ $b$.test(bmer)}
  				{$q$.append(i)}
  				}
  			}
  return q 
}

\Fn{\ALIGNMENT{g, q}}
{
return bwa(g, q)\\
}

\Fn{\MERGE{M}}
{
$S$ = \merge{M} \tcp*[h]{call DIDA merge}\\
return S
}
\vspace{-1ex}
\end{algorithm}

However, the input reads need to be protected. Together with the partitioned reference genome, inputs are fed into the dispatch process to get the same number of dispatched reads as the partitioned reference genome. The partitioned reference genome and the dispatched reads are then distributed to a cluster of nodes for the parallel running of the actual alignment (or run sequentially on one node). The partial results are then merged to form the final output, which will also be encrypted.

In Figs. ~\ref{fig:overall_overview}-\ref{fig:merge}, all the processes running within SGX are depicted in blue boxes. The alignment process in the worker nodes uses Scone to minimize the source code revision, which is required to handle sensitive data securely. The input and output SAM files are stored in a protected folder. They are handled transparently by the Scone file protection feature \cite{scone-fspf}. All nodes are from the same cluster and have access to a common shared file system where the intermediate results between processes are all encrypted. Each process within an SGX is undergoing the unsealing/sealing process to securely read the data and write the output to the file system.

% --------------------------------

\subsection{Trust Establishment}\label{subsec:keyexchange}

When the data owner wants to delegate a job to HySec-Flow, the owner needs to know that the service provider truly provides the service on a trusted platform. Therefore, the owner can initiate remote attestation to establish mutual trust. Since the source code of HySec-Flow is public, the data owner can easily know whether the remote service is running in a trusted control node enclave or not through verifying the measurement, which can be derived from the enclave source code. The RA-TLS protocol can be integrated into our work for trust establishment and key exchange. After mutual trust between the data owner and service provider is established via remote attestation, a key $K_D$ can be generated by ECDH key exchange to securely communicate and transfer data. 
It should be noted that the key agreement step can be done using the attestation feature of Scone's premium version or using Graphene-SGX's remote attestation plan, so we don't implement it by ourselves.

The data owner can then transfer data files encrypted using this key, and these files can then be decrypted in the work nodes' enclaves at the server-side. Notice that the enclaves between the control node and work nodes also need to establish mutual trust, and the key $K_D$ to decrypt data files should also be passed through a secure channel. Yet intermediate data files can be securely stored in untrusted storage, such as in a shared file system, and be transferred via an untrusted channel since they are encrypted.  Finally, the framework can encrypt and return the result to the data owner.

% --------------------------------

\subsection{Partition}

This stage is performed to split the reference genome sequence into multiple partitions such that each partition can be individually indexed and searched on different nodes of the cluster. The partitioner takes the reference genome as the input and outputs p number of partitions as shown in steps 1 and 2 of Fig. \ref{fig:overall_overview}. The partition task works only on non-sensitive data and can run on a single node for a simple pass through the reference genome sequence. For the same p and same reference genome, partitioning will only execute once.

\subsection{Indexing}

The partitions generated are indexed using a popular read alignment tool like BWA. This operation can be performed parallelly on each partition utilizing the available computing resources of the cluster. Furthermore, this operation does not require to be running in a secure environment. Hence this step reads and writes to the non-secure shared file system as shown in steps 3 and 4 of Fig. \ref{fig:overall_overview}.

\subsection{Dispatch}

The dispatch stage is performed to \textbf{reduce the search space} of an input DNA sequence within each partition. This can be performed by utilizing many application-dependent techniques. We adapt an approach based on the bloom filters from DIDA. We compute a bloom filter for each partition by inserting sub-sequences of length 'b' of the reference genome partition with overlaps of length 'l'. Bloom filter generation works only on non-sensitive data. We perform this part of the dispatch process entirely outside Intel SGX (step 5 of Fig. \ref{fig:overall_overview}). Furthermore, generated bloomfilters can be reused for future executions as long as 'b' and 'l' remain the same. Hence we persist generated bloom-filters to the disk as a binary file(6 of Fig. \ref{fig:overall_overview}). Bloom-filters generated per each partition will be assigned with a uniquely identifiable name generated based on the reference genome and the 'b' and 'l' arguments. Having a separate binary file for each bloomfilter makes running dispatch inside the limited enclave memory efficient.

\begin{figure}[t]
\vspace{-1ex}
  \centering
    \includegraphics[width=0.8\linewidth,height = 0.52\linewidth]{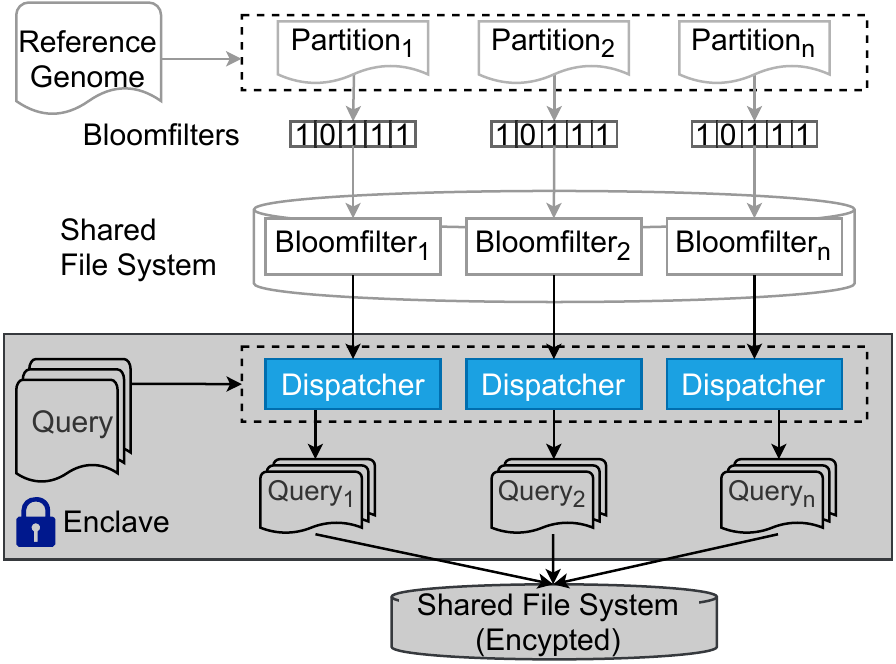}
  \caption{Dispatch}
\label{fig:dispatch}
\vspace{-4ex}
\end{figure}

We assume input DNA sequences are in the encrypted form when we receive them into the framework. The next stage of the \emph{dispatch} task is looking up the bloom-filters to determine the membership of the subsequences of the input DNA sequence within each partition. Since dispatch involves sensitive data, we have modified the DIDA framework to execute the bloom filter lookup logic inside the SGX enclaves. Input partitioning is performed by first loading the encrypted DNA sequence into the SGX enclave and then decrypting internally to extract the unencrypted data. Then we create empty string builders within the enclave (Line 6 of algorithm \ref{alg:functions}) to hold the input sequence for each partition. Finally, bloom filter lookups are performed to determine the membership. In case of a positive lookup in the bloom filter, we append the input subsequence to the corresponding string builder. As shown in Fig. \ref{fig:bloomfilter} and explained in section \ref{sec:bloomfilter}, we expect false positive responses for some of the lookups. But overall, this approach reduces the search space for the alignment step significantly. Furthermore, the false-positive rate can be controlled by configuring the size of the bloom-filter. We then persist input partitions into the disk by encrypting the files transparently using the file protection features provided by Scone or Graphene.

The dispatch step can be parallelly run for each reference genome partition as depicted in line 4-5 of Algorithm \ref{alg:functions}. The output will be saved back to the protected file system as shown in step 9 of Fig. \ref{fig:overall_overview}.

\subsection{Alignment}

Together with the corresponding index of the partitioned reference genome, the dispatched reads file is assigned to the cluster's worker nodes for the alignment process (Step 10 of Fig. \ref{fig:overall_overview}). This step could run sequentially on a single node or distribute over multiple nodes. As a proof-of-concept, we use BWA for the actual alignment of the reads with the Scone framework to leverage the SGX capability. Minor changes on the BWA code are needed so it works with the Scone file protection \cite{scone-fspf} setup. It provides transparent file decryption and encryption of the input and output files for the alignment setup. The code is compiled within a docker image from Scone that provides a cross compiler toolset and run in a docker container. This approach is generally applicable to other legacy applications, like BWA, to run within SGX. While using the BWA tool for this step, other alternative tools could be used, or even customized programs developed totally with SGX SDK, in which case Scone would not be needed anymore.
\begin{figure}[ht]
\vspace{-1ex}
  \centering
\includegraphics[width=0.8\linewidth,height=0.46\linewidth]{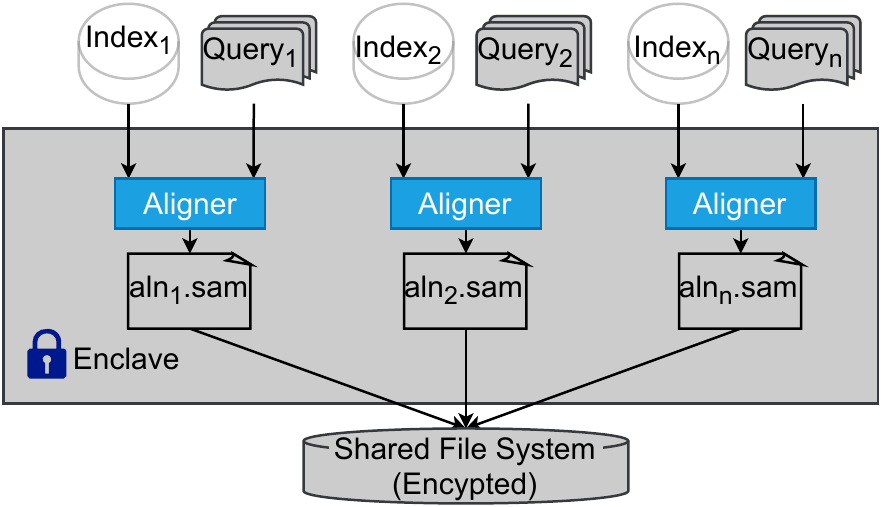}
  \caption{Alignment}
\label{fig:alignment}
\vspace{-1ex}
\end{figure}
The results of this step are the partial SAM output (Step 11 of Fig. \ref{fig:overall_overview}) from each dispatched reads and partitioned reference. Once all the results are ready, they will be merged in the following step to form the final results. In the evaluation experiment, we considered the input reads files containing either the paired-end reads, in which two reads were sequenced from a short distance (i.e., 300-500 base pairs apart) in the genome, or the single-end reads each read was sequenced independently. Thus, the reads alignment task for these two types of input is referred to as the paired-end alignment and the single-end alignment respectively.

\subsection{Merge}
This task expects multiple encrypted SAM files (Step 12 of Fig. \ref{fig:overall_overview}) as the input and performs merging techniques in the DIDA framework inside the SGX enclaves. The encrypted input SAM files will be decrypted only within the SGX enclave. Once the merging is done, the output will be sealed (Step 13 of Fig. \ref{fig:overall_overview}) using the user's shared key since this is the final output expected by the user. Besides sealing the final output and unsealing the initial input, we have delegated encryption and decryption of the intermediate inputs and outputs to the transparent file system's encryption mechanisms provided by Scone or Graphene. 

\begin{figure}[ht]
\vspace{-2ex}
\centering
\includegraphics[width=0.65\linewidth,height=0.25\linewidth]{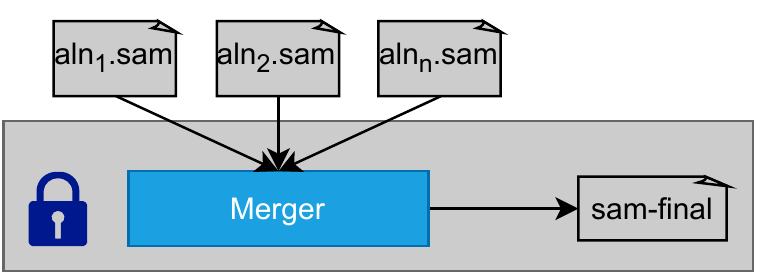}\vspace{0.1cm}

 \caption{Merge encrypted SAM files}
\label{fig:merge}
\vspace{-3ex}
\end{figure}

\subsection{Pipeline}

Algorithm \ref{alg:Pipeline} shows how we can run the above stages in a distributed pipeline to leverage the resources available across the cluster. For a given user query, the first task scheduled will generate the bloomfilters. If the bloomfilters are already available in the disk for the provided arguments, this stage completes immediately. The next task is to run the dispatch in a secure environment. So the resource scheduler is configured to schedule dispatch tasks into nodes having SGX hardware capabilities. Once the dispatched task is completed, the alignment tasks can be parallelly scheduled on multiple SGX nodes as shown in Algorithm \ref{alg:Pipeline}, lines 7-8. Once all the dispatch tasks are completed, the merge can be scheduled on a secure node to generate the final output.

\subsection{Data Sealing}\label{subsec: datasealing}
All information that lies outside the trusted parts (enclave) in the workflow should be in ciphertext state. Therefore we propose sealing/unsealing modules inside the enclave to encrypt/decrypt intermediate data across nodes.

Assuming the remote attestation has been done before the data owner's input is uploaded to the framework, a session key can be retrieved to establish a secure channel between the genomic data owner and the framework. HySec-Flow can accept file input in plaintext and can do the initial encryption for the user. Besides, to protect the data transferring between enclaves from the outside attacker, we seal the output data and unseal the input data with the same key. To this end, secure channels can be built.

\section{Evaluation}
%In this section, we report our security analysis and performance evaluation.
\subsection{Security Analysis}

SGX Enclave can protect the code/data integrity even when the executable is loaded into a library OS (e.g., Graphene-SGX can provide a measurement hash against the loaded code/library for checking).
Moreover, disk I/O has been safeguarded by Scone/Graphene's protected filesystem, which utilizes AES-GCM to encrypt user data and immediate data during the computation. Under our threat model, the only security risk is key delivery, which is protected by the secure channels we built after trust establishment. Therefore, file tampering attacks can be defeated.

Side channels have been considered to be a threat to trusted execution environments, including SGX. There is a line of research that identifies such security risks~\cite{lee2017inferring,wang2017leaky,van2018foreshadow,chen2019sgxpectre}. In the meantime, prior research also shows that most of the side channel risks can be detected or mitigated using certain defense strategies~\cite{shinde2016preventing,shih2017t,oleksenko2018varys,sinha2017compiler,chen2018racing}. 
Most prior studies on SGX-based computing systems consider side channels to be outside their threat models~\cite{arnautov2016scone,tsai2017graphene,shen2020occlum} with the continuous interest in the topic, as Intel assumes when developing the TEE~\cite{van2018tutorial}. 
Our research follows this threat model and has not implemented known protection (including those against a set of micro-architectural threats) in our prototype. In future research, we will evaluate the impacts of side-channel leaks on our computing frameworks and genomic data analysis tasks and build into our system a proper protection when necessary.

\subsection{Experimental setup \& data sets}
Our experiments are conducted on a 10-nodes SGX-enabled cluster, with each node has an Intel(R) Xeon(R) CPU E3-1280 v5 @ 3.70GHz CPU and 64G RAM. The SGX enclaves are initialized with 8GB heap space with both Scone and Graphene. The libraries are ported into Graphene include \textit{ld.so}, \textit{libc.so}, \textit{libm.so}, \textit{libdl.so}, \textit{libutil.so}, and \textit{librt.so}. We also port \textit{libpthread.so} for multi-threading support. Scone containers are based on Scone's alpine linux images running Scone version 1. 
We use datasets from the 1000 Genome project \cite{siva20081000} for the testing. Without loss of generality, for single-end alignment, we use the SRR062634.filt.fastq, which has $\sim$309K reads, with 100bp per read. For paired-end alignment, we use SRR062634\_1 and SRR062634\_2. These files are arbitrarily selected as a personal genome. The detailed data set and specification are shown in Table \ref{tab:data}.

\begin{table}[ht]
\vspace{-1ex}
\centering
\small
\caption{Dataset specification.}
\begin{tabular}{l|l|r|r}

\thickhline
&&&\\[-0.25cm]
\bf{Data Set}&  \bf{Source}   & \bf{\# Reads}& \bf{bp/read}\\[0.1cm]
\hline
&&&\\[-0.2cm]
SRR062634.filt.fastq   & 1000 Genomes\cite{siva20081000}   &309K          & 100 \\
 SRR062634\_1   &  1000 Genomes\cite{siva20081000} &24M         &100   \\
SRR062634\_2  &  1000 Genomes\cite{siva20081000} &24M          &100  \\[0.05cm]
\thickhline

\end{tabular}
\label{tab:data}
\vspace{-3ex}
\end{table}

\subsection{Accuracy}
Although the bloomfilter-based dispatch step narrows down the search space for subsequent steps greatly, that comes with an impact on the accuracy of the final output. However, the scope of our approach is to perform reads mapping with an acceptable accuracy securely. Hence we consider the accuracy of the final outputs from DIDA's approach as the baseline. We compare the output files generated by the merge stage after running dispatch, alignment, and merge in sequence on Scone and outside Scone. When Scone outputs are decrypted to obtain the plain text output, it matches exactly with the output from the non-Scone execution.

\subsection{Benchmark of SGX overhead}
Using SGX could introduce overhead from multiple aspects. 
\subsubsection{Overhead from enclave initialization}
Enclave initialization overhead is impacted by the heap size requested. We measure the enclave initialization time by varying the HeapMaxSize (16M, 64M, 256M, 1024M, 4096M). The results show a good linear relationship with the increasing max size of heap/stack. We observe that enclave initialization time is about 0.04 seconds per MB of the configured maximum heap size.

When developing an SGX application using SGX SDK in enclave configuration file Enclave.config.xml, we can set the parameters StackMaxSize and HeapMaxSize. These parameters determine the estimated memory requirements of the generated enclave.

\subsubsection{Overhead from OCall/ECall}
The SGX-enabled program defines an interface using Enclave Definition Language (EDL), in which ECalls and OCalls are defined. A program can only invoke these defined methods to cross the untrusted and trusted execution environment boundary. We measured the overhead of the invocation of these calls. The overhead of OCall and Ecall are 5.27 and 4.65 seconds per million calls respectively. As a comparison, making the same calls within the untrusted environment only costs 1.3 ms per million calls. 

\subsubsection{Overhead from EPC page swapping}
An enclave can only utilize what a Processor Reserved Memory (PRM) can provide at the current stage, which is 128MB. In actual use, the usable memory size for an SGX application is only around 90MB, and the system uses the rest. Therefore, enclave Page Cache (EPC) can only use this memory. 
When a larger dataset may not fit into this space, an EPC page swap occurs, and this process introduces high overhead. For example, for data access pattern within the memory region of size $\sim$40MB, the results show that 1 billion runs of the emulated code block, when very few page faults occurred ($\sim 10^4 $), the execution time is around 3 seconds. However, when we need to frequently access data outside of that region and thus EPC page swap occurred more frequently (more than $10^7$ times), the execution time is around 300 seconds, which is about 100 times slower.

\subsection{Optimal partitions for splitting the reference genome}

\begin{figure}[ht]
\vspace{-2ex}
  \centering
    \includegraphics[width=0.85\linewidth]{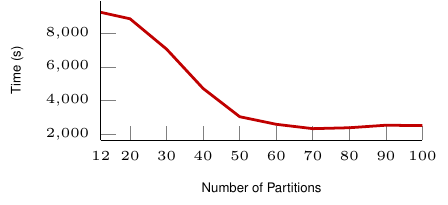}
  \caption{Sequential run time within SGX for different number of partitioned reference genomes.}\label{fig:seqtime-vs-nsplits}
  \vspace{-1ex}
\end{figure}

We have experimented with a different number of partitions for the reference genome to find the optimal configuration. Fig. \ref{fig:seqtime-vs-nsplits} shows the results. The runtime is measured by sequentially run the alignment for dispatched reads on one single node using SGX via Scone. We notice that with the increasing number of partitions, the overall runtime decrease. However, when the number of partitions is greater than 60, it got flattened. Considering the human reference genome data we used is about 3.2 GB, this translates to the reference partition size around or smaller than 50 MB. With the usable memory space around 90 MB for SGX, this optimal configuration suggests that the entire indexing table can fit into the SGX EPC to minimize the unnecessary EPC swapping, thus improving the overall performance. We use reference genome partition number 80 as the optimal number to run within SGX in our future experiments.

\subsection{Execution times of dispatch and merge}
\begin{table*}[ht]
\begin{center}
\small
\caption{Dispatch, Alignment and Merge for BWA (Single End Reads)}
\begin{tabular}{r|rrr|rrr| rrr|rrr| rr}
\thickhline
\multirow{3}{*}{}   & \multicolumn{6}{c|}{} &\multicolumn{6}{c|}{} & \multicolumn{2}{c}{} \\[-0.3cm]

\multirow{3}{*}{\textbf{\# Partitions}}   & \multicolumn{6}{c|}{\textbf{Dispatch} (seconds)} &\multicolumn{6}{c|}{\textbf{Alignment} (seconds)} & \multicolumn{2}{c}{\textbf{Merge} (seconds)} \\
 & \multicolumn{3}{c}{\textbf{Non Secure}} & \multicolumn{3}{c|}{\textbf{Secure}} &\multicolumn{3}{c}{\textbf{Non Secure}} & \multicolumn{3}{c|}{\textbf{Secure}}& \multicolumn{1}{l}{\textbf{Non Secure}} & \multicolumn{1}{l}{\textbf{Secure}} \\
\cline{2-13}
\textbf{}  & \multicolumn{1}{c}{\textbf{min}} & \multicolumn{1}{c}{\textbf{avg}} & \multicolumn{1}{c|}{\textbf{max}} & \multicolumn{1}{c}{\textbf{min}} & \multicolumn{1}{c}{\textbf{avg}} & \multicolumn{1}{c|}{\textbf{max}} & \bf min&\bf avg&\multicolumn{1}{c|}{\textbf{max}}& \bf min&\bf avg&\bf max& \multicolumn{1}{l}{} & \multicolumn{1}{l}{} \\[0.1cm]
\hline
10    & 14.45 & 14.52 & 14.45 & 44.00 & 44.58 & 44.87 & {0.27} & {0.43} & {0.77} & {5.19} & {7.58} & {8.70} &0.88 & 4.70\\
20    & 6.57 & 7.81 & 8.06 & 22.35 & 26.62 & 27.26 & {0.12} & {0.14} & {0.16} & {2.32} & {4.10} & {4.77} &0.83 & 4.65 \\
40    & 3.20 & 4.46 & 4.56 & 9.63 & 14.32 & 14.58 &  {0.06} & {0.10} & {0.14} & {1.05} & {1.77} & {2.15} &0.80 & 4.62 \\
80    & 1.48 & 1.48 & 2.93 & 6.29 & 9.12 &\bf 9.68 & {0.03} & {0.13} & {0.47} & {0.44} & {0.73} & \bf{1.18} &0.81 & \bf4.66 \\
\thickhline
\end{tabular}\\
\label{tbl:dispatch_merge_sgx}
\end{center}
\raggedright{\small The alignment step is parallelly executed across the cluster. Minimum, Average and Maximum time reported by containers.}

\normalsize
~\\
Table \ref{tbl:dispatch_merge_sgx} shows the results for the proposed partition and dispatch. The partition and dispatch approach shows a slowdown which greatly improves for higher partition counts due to better EPC utilization. Our approach makes it easier to run in parallel because of the pleasingly parallel nature of the data and the application. In the best case(if resources are available), we can run the single end alignment pipeline securely in 15.53 seconds (based on 9.68s in parallel dispatching, 1.18s in parallel alignment over 80 nodes and 4.66s in merging) by partitioning the problem into 80 subtasks. Even in the worst case that has only one SGX enabled node, we can expect to complete the alignment in 792.46 seconds running sequentially.
\vspace{-2ex}
\end{table*}

\begin{table}[ht]
\centering
\small
\caption{ Non Secure Execution (one-time calculations)}
\begin{tabular}{r|r|r|r}

\thickhline
&&&\\[-0.25cm] 
\bf{\#Partitions}&  \textbf{Partitioning}  (s)  & \makecell[l]{\bf{Bloomfilter}\\ \textbf {Building}\\ (seconds) }  & \makecell[l]{\textbf{Indexing} \\ (seconds)} \\[0.4cm]
\hline
&&&\\[-0.25cm]
1  &0    &1985.09         &4302 \\
10  &37.59     &1211.65          &3052 \\
20  &39.39    &1113.9     &2853   \\
40 &40.17   &1147.48        &2602  \\
80 &41.93    &1316      &2292  \\
\thickhline

\end{tabular}
\vspace{-3ex}
\end{table}

Table \ref{tbl:dispatch_merge_sgx} lists the time used for the dispatch and merge steps (single-end reads). Although bloom filter building seems to be dominating the entire workflow, it is a one-time operation. The same set of bloomfilters can be used for subsequent executions on different user inputs. Also, we notice that partition size does not significantly impact the execution time as these two stages of the workflow are not parallelized. However, the dispatch step can be parallelized to run in parallel on Bf\textsubscript{n} and Query to produce corresponding Query\textsubscript{n} in contrast to the Dispatch step shown in Fig \ref{fig:dispatch}. If computing resources are available, this should reduce the execution time by a factor of 'p', where 'p' is the number of partitions. 
\vspace{-0.6ex}

\subsection{Execution times of Data Sealing/Unsealing}

We use RDTSC (returns a 64-bit time stamp counter (TSC), which is increased on every clock cycle) to measure the time consumption outside the enclave of data sealing/unsealing functions we built. Each test runs 10 times. As for sealing inside, we implement an OCall for timing. The OCall checks the outside TSC value and itself costs less than 0.01ms. 
Table~\ref{tab:sealing} shows the average execution time when different datasets are given. When dealing with the single-end input SRR062634, the sealing time is less than 3s. 
For the pair-ended data (SRR062634\_1 and SRR062634\_2), the sealing time is less than 10s. 

\begin{table}[ht]
\vspace{-0.5ex}
\begin{center}
\small
\caption{Data Sealing/Unsealing with Intel SGX}
\begin{tabular}{m{20ex}|m{15ex}m{15ex}}
\thickhline
\\[-0.3cm]
\textbf{Operation}         & \makecell[c]{\textbf{Single End} \\ (seconds)} &  \makecell[c]{\textbf{Pair End} \\ (seconds)}\\[0.1cm]
\hline
\\[-0.3cm]
Sealing outside         & \makecell[c]{2.59}                          & \makecell[c]{8.29}   \\ 
Unsealing inside & \makecell[c]{2.59}                         & \makecell[c]{8.30}      \\ 
Sealing inside & \makecell[c]{2.59}                                & \makecell[c]{8.31 }          \\

\thickhline
\end{tabular}\label{tab:sealing}
\end{center}
\vspace{-5ex}
\end{table}

\subsection{Execution Time of Reads Mapping}
Table \ref{tbl:exec_single_ref}  lists the execution time for the reads mapping tasks in different settings, which includes single-end and paired-end execution times. The overhead of using SGX and the speedup of our proposed solution is compared to the direct Scone and Graphene solutions. The experiment setup and scripts can be found at \cite{dsc-sgx-bwa}, \cite{graphene-bwa} and \cite{code}. 

\begin{table}[t]

\small
\centering
\caption{BWA Alignment (Sequential)}
\label{tbl:exec_single_ref}
\begin{tabular}{l|lrr} 
\thickhline
&&&\\[-0.3cm] 
\textbf{Alignment Type}& \makecell[l]{\bf{Containers}} &\makecell[l]{ \textbf{BWA}\\ \textbf{Alignment}\\(seconds)} & \makecell[l]{ \bf{Slowdown}}\\[0.4cm]
\hline
&&&\\[-0.3cm] 
\multirow{3}{*}{Single-end }&Non  Secure & 91 & 1\\ 
 &SCONE   & 3291  & {\textbf{36.1}} \\ 
 &Graphene      & 10603 & {\textbf{\hspace{0.55ex}117}} \\[0.05cm]
\hline
&&&\\[-0.3cm] 
\multirow{3}{*}{Paired-end }& Non  Secure  & 15423 &1 \\ 
 & SCONE  & $>$173K    & {\textbf{$>$41}} \\ 
& Graphene       &$>$173K & {\textbf{$>$41}}   \\
\thickhline

\end{tabular}

\raggedright{ ~\\A single-end reads file has 309K reads and each read is 100bp long. A pair end of reads file has 24M reads each. Non secure refers to BWA execution w/o SGX.}
\vspace{-4ex}
\end{table}

\begin{table*}[ht]
\begin{center}
\small
\caption{Comparison of HySec-Flow against Scone or Graphene for Running BWA}
\label{tab:total_time}
\begin{tabular}{r|rrr|rrr}
\thickhline
&&&&\\[-0.3cm]
\makecell[l]{\textbf{Number of Partitions}}  & \makecell[c]{\textbf{SCONE}\\ \footnotesize (Sequential)\\\footnotesize Total time (s)} &\makecell[c]{\textbf{HySec-Flow}\\ \footnotesize parallel SCONE\\\footnotesize Total time (s)}  &  \makecell[c]{\textbf{Speedup}} &\makecell[c]{\textbf{Graphene}\\ \footnotesize (Sequential)\\\footnotesize Total time (s)} &\makecell[c]{\textbf{HySec-Flow}\\\footnotesize parallel Graphene\\\footnotesize Total time (s)}&  \makecell[c]{\textbf{Speedup}} \\[0.05cm]
\hline
&&&&\\[-0.3cm]
1 & 3291&3227.48  & \bf{1.02} &10603  &11025.41  &0.96  \\ 
10   & 3291      & 58.27 & \bf{56.47}&10603 &412.66 &\bf 25.7  \\ 
20& 3291 & 36.68 & \bf 89.71&10603 & 300.9   &\bf 35.23  \\ 
40& 3291& 21.34 &\bf 154.19 &10603 & 245.84  &\bf 43.13     \\
80& 3291& {15.52} &{\bf 211.94}&10603 &{217.23}    &{\bf 48.80}    \\
\thickhline
\end{tabular}
\end{center}
\vspace{-4ex}
\end{table*}

\begin{figure*}[ht]
\centering
\includegraphics[width=0.35\linewidth]{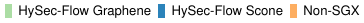}
\vfill
\captionsetup[subfigure]{oneside,margin={1.4cm,0cm}}
\subfloat[\bf Total Secure Execution Time\\ (Dispatch, Alignment and Merge)]{\includegraphics[width=0.3\linewidth]{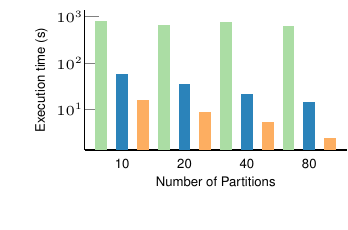}}
\hspace{1ex}
\subfloat[\textbf{Dispatch (Sequential)} ]{\includegraphics[width=0.3\linewidth]{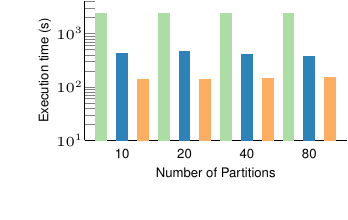}}
\hspace{1ex}
\subfloat[\textbf{Dispatch (Parallel)} ]{\includegraphics[width=0.3\linewidth]{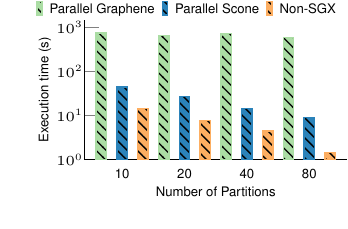}}
\vfill
\captionsetup[subfigure]{oneside,margin={1.4cm,0cm}}
\subfloat[\bf Non-Secure Execution Time\\ (Partitioning, Indexing, and Bloom Filter)]{\includegraphics[width=0.3\linewidth]{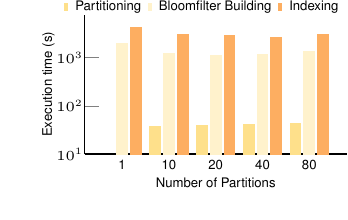}}
\hspace{1ex}
\subfloat[\textbf{Alignment} ]{\includegraphics[width=0.3\linewidth]{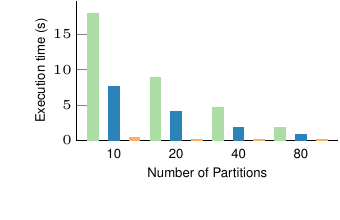}}
\hspace{1ex}
\subfloat[\textbf{Merge} ]{\includegraphics[width=0.3\linewidth]{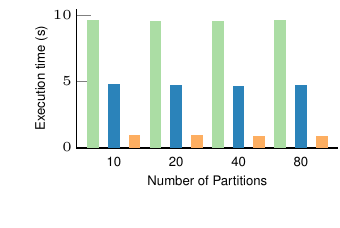}}
\caption{Comparison of the HySec-Flow execution time of Scone and Graphene in different stages.}
\vspace{-3ex}
\end{figure*}
The file protection features of both Scone and Graphene are configured and enabled, so the input fastq files and output SAM files are secured. The overhead is measured against the non-SGX approach from Table \ref{tbl:exec_single_ref}. The speedup can be determined against the SGX Scone solution from the same table.

As shown in Table \ref{tbl:exec_single_ref}, running BWA directly without involving SGX was fast, but data privacy is not protected. Another approach is running BWA within a Scone container. This approach provides an easy way to utilize SGX, but the performance penalty is huge due to the EPC size limitation of SGX and the frequent page swapping when dealing with big data. We observe about 40x-50x slowdown comparing to the non-SGX setup. Graphene-SGX performs worse than Scone because it causes more paging overhead when loading more components with the whole LibOS to the enclave. 

\begin{figure}[ht]
  \vspace{-1ex}
  \centering
\includegraphics[width=0.85\linewidth]{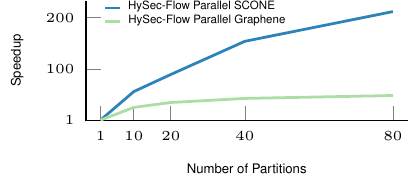}
  \caption{Speedup of HySec-Flow over Scone and Graphene}
  \label{fig:time_projection}
  \vspace{-2ex}
\end{figure}

Although the bloom filter generation is mostly a one-time operation, if we consider that time into account too, the execution times will increase only by 1316 seconds. \textbf{The best-case of HySec-Flow execution time (15.52 seconds) is 6x and 212x speedup compared to non-SGX execution (91 seconds) and Scone execution (3291 seconds) respectively.} The total execution times and projected variation of speedups for other parallelism configurations(10, 20, 40, 80) are shown in Table \ref{tab:total_time} and illustrated in Fig. \ref{fig:time_projection}.

\section{Related Work}
DIDA \cite{mohamadi2015dida} is a distributed indexing-dispatched alignment framework for reads mapping. Our approach is inspired by the DIDA framework but has taken data privacy into full consideration: the computation involving sensitive data is executed in the SGX enclave, and these sensitive data remained encrypted outside the enclave. We use customized data and computation partitions to split human reference genome sequence into small segments so that each reads mapping subtask does not consume much memory. This produces better performance running inside an enclave. In contrast, the original DIDA framework only supports a small number of subtasks partitions, each comprising a long reference genome sequence of the whole chromosome.

Scone \cite{arnautov2016scone} provides an easy-to-use container environment that can run executable code within the SGX enclave.
We use Scone to run the individual alignment {\em worker} program in HySec-Flow. However, executing codes in enclaves directly using Scone alone may introduce significant performance overhead due to the lack of optimization on the data access according to the limited SGX EPC space. Our proposed framework addresses this issue by splitting data into smaller segments and running multiple jobs sequentially in a single enclave or parallel in multiple enclaves.

Graphene-SGX is a practical library OS for unmodified code on SGX. It uses Graphene LibOS \cite{tsai2014cooperation} as the inner core to support the binary code compatibility \cite{shanker2020evaluation}. The enclave consists of the application to be protected linked with a library OS. Graphene-SGX can execute the applications by writing a manifest file that describes (among other things) the set of libraries used by the benchmark (among other things). Compared to Scone, Graphene can provide a more flexible configuration of multithreading support.

Although existing SGX-based secure computing approaches often assume side channels as an orthogonal research topic~\cite{sinha2015moat,subramanyan2017formal,shen2020occlum}, side channels impose serious threats to secure computing using SGX as attackers can use them to circumvent explicit security defenses implemented by SGX. A rich literature has focused on discovering SGX side channels~\cite{lee2017inferring,wang2017leaky,van2018foreshadow,chen2019sgxpectre}. Notably, HySec-Flow is also vulnerable to such threats. Fortunately, most known side channels in SGX-based computation can be detected or mitigated using various defense strategies~\cite{shinde2016preventing,shih2017t,oleksenko2018varys,sinha2017compiler,chen2018racing}. 

\section{Conclusion}
We have introduced an architecture for an end-to-end workflow of privacy-preserving genomic data analytics using Intel's SGX technology. We use the reads mapping application (specifically the commonly used BWA algorithm) to showcase the usability and the performance of the framework. The naive Scone solution has modest performance improvement on single-node even when using the partition and dispatch methods. \textbf{HySec-Flow makes it possible to run in parallel on multiple nodes while still in a secured fashion. 
When tested with single-end reads mapping tasks, we've observed a speedup of up to 212x (for 80 partitions) compared to the naive approach directly executing BWA within the Scone framework. The speedup is mainly achieved from the process level parallelism as well as significantly reduced search space by the bloomfilter based dispatch step.}

We stress that HySec-Flow can be easily adapted to a category of many genomics applications where the algorithms are pleasantly data-parallel, e.g., for genome variation calling ~\cite{lambert2018maskal,mandal2018data}, 
for gene expression analysis using RNA-seq data~\cite{prasad2016global}, and peptide identification in clinical proteomics~\cite{decramer2008urine}. However, in each of these cases, we need to devise a customized data partition algorithm that can assemble subsets of input data for subtasks so that the subtasks are performed most efficiently.
\vspace{-0.5ex}
\section{Future Work}

The HySec-Flow framework can be extended to handle multiple search tasks from different users by adding a new 'driver' component to securely accept jobs from users and assign containers on demand from a heterogeneous pool of containers due to the pleasingly parallel nature of the workloads.

We will further integrate into future work another sophisticated framework, Harp \cite{ zhang2015harp,zhang2016high,chen2017benchmarking,peng2017harplda+,peng2019harpgbdt,zhang2016model,chen2019subgraph2vec,peng2021rank}, which utilizes MPI-style collective communications to deal with Big Data among the nodes from a cluster in an HPC-Cloud environment with an SGX-enabled machine learning applications.

The HySec-Flow framework has been designed to support non-secure tasks, secure tasks written directly on Intel SGX API, and secure tasks on Scone or Graphene. Hence other hybrid workflows (secure / non-secure) other than genome sequencing can be ported into the framework and scale infinitely using a programmable API\cite{code_framework}. Reads mapping is a large data-intensive computing task compared to previously developed SGX-based solutions (e.g., variant searching and GWAS). Therefore, the framework presented here can be extended to implement privacy-preserving algorithms for other data-intensive genome computing tasks such as genome variation calling ~\cite{10002010map} and gene expression analyses ~\cite{alarcon2008linkage} in future work. 
\vspace{-1ex}
\section{Acknowledgment}
This work is partially supported by NSF grant No.1838083 on BIGDATA: IA: Enabling Large-Scale, Privacy-Preserving Genomic Computing with a Hardware-Assisted Secure Big-Data Analytics Framework, NSF grant CCF-1918626 Expeditions: Collaborative Research: Global Pervasive Computational Epidemiology,  NSF grant No. 1835631 CINES: A Scalable Cyberinfrastructure for Sustained Innovation in Network Engineering and Science, and NIH R01HG010798: Secure and Privacy-preserving Genome-wide and Phenome-wide Association Studies via Intel Software Guard Extensions (SGX). We appreciate technical support from Intel Inc. and would like to thank Robert Henderson and the system team for their assistance with our experiments on the SGX cluster.

\small
\vspace{-0.5ex}
\bibliographystyle{./bibliography/IEEEtran}
\bibliography{./bibliography/IEEEabrv,./bibliography/ref}

\end{document}